\documentclass{article}

\usepackage{microtype}
\usepackage{graphicx}
\usepackage{booktabs} 
\usepackage[T1]{fontenc}
\usepackage[utf8]{inputenc}
\usepackage{tabularx}
\usepackage{multirow}
\usepackage{array}
\usepackage{url}
\usepackage{hyperref}
\usepackage{enumitem}
\usepackage{amsmath}
\usepackage{xspace}
\usepackage{listings}
\usepackage{algpseudocode}
\usepackage{balance}
\usepackage[linesnumbered,ruled,vlined]{algorithm2e}
\usepackage{subcaption} 
\usepackage{amssymb}

\SetKwInOut{Require}{Require}
\SetKwInOut{Ensure}{Ensure}
\SetKwProg{Proc}{Procedure}{}{}
\SetKwProg{Fn}{Function}{}{}
\SetKwProg{try}{try}{}{}
\SetKwProg{catch}{catch}{:}{end}
\SetKwProg{finally}{finally}{:}{end}
\SetKwFunction{ExtractPayload}{ExtractPayload}
\SetKwFunction{CreateTask}{CreateTask}
\SetKwFunction{Enqueue}{Enqueue}
\SetKwFunction{CloneOrUpdateRepo}{CloneOrUpdateRepo}
\SetKwFunction{GetBuildConfig}{GetBuildConfig}
\SetKwFunction{RunStaticAnalysis}{RunStaticAnalysis}
\SetKwFunction{GetDiffChunks}{GetDiffChunks}
\SetKwFunction{GroupViolationsByChunk}{GroupViolationsByChunk}
\SetKwFunction{ExtractCodeSnippet}{ExtractCodeSnippet}
\SetKwFunction{GetCodeContext}{GetCodeContext}
\SetKwFunction{GenPrompt}{GenPrompt}
\SetKwFunction{QueryLLM}{QueryLLM}
\SetKwFunction{AppendToReport}{AppendToReport}
\SetKwFunction{FormatPRComment}{FormatPRComment}
\SetKwFunction{PostCommentToPR}{PostCommentToPR}
\SetKwFunction{Cleanup}{Cleanup}
\SetKwFunction{GenerateKey}{GenerateKey}
\SetKwFunction{CreatePayload}{CreatePayload}
\SetKwFunction{GeneratePrompt}{GeneratePrompt}
\SetKwFunction{Format}{Format}
\SetKwFunction{ProcessLLM}{ProcessLLM}
\SetKwFunction{FormatError}{FormatError}

\newcolumntype{Y}{>{\raggedright\arraybackslash}X}

\lstdefinestyle{prompt}{
  basicstyle=\ttfamily\footnotesize,
  columns=fullflexible,
  frame=single,
  breaklines=true,
  keepspaces=true
}

\usepackage[accepted]{mlsys2025}

\mlsystitlerunning{Grounded AI for Code Review}

\begin{document}

\twocolumn[
\mlsystitle{Grounded AI for Code Review: Resource-Efficient Large-Model Serving in Enterprise Pipelines}




\begin{mlsysauthorlist}
\mlsysauthor{Sayan Mandal}{to}
\mlsysauthor{Hua Jiang}{to}
\end{mlsysauthorlist}

\mlsysaffiliation{to}{AMD, San Jose, CA, USA}

\mlsyscorrespondingauthor{Sayan Mandal}{sayan.mandal@amd.com}
\mlsyscorrespondingauthor{Hua Jiang}{hua.jiang@amd.com}

\mlsyskeywords{Grounded AI, Automated Code Review, Resource-Efficient LLM Serving, CI/CD Integration, Hybrid Static Analysis}

\vskip 0.3in

\begin{abstract}
Automated code review adoption lags in compliance-heavy settings, where static analyzers produce high-volume, low-rationale outputs, and naive LLM use risks hallucination and incurring cost overhead. We present a production system for grounded, PR‑native review that pairs static-analysis findings with AST-guided context extraction and a single-GPU, on-demand serving stack (quantized open-weight model, multi-tier caching) to deliver concise explanations and remediation guidance. Evaluated on safety-oriented C/C++ standards, the approach achieves sub-minute median first-feedback (offline p50 build+LLM 59.8s) while maintaining competitive violation reduction and lower violation rates versus larger proprietary models. The architecture is decoupled: teams can adopt the grounding/prompting layer or the serving layer independently. A small internal survey (n=8) provides directional signals of reduced triage effort and moderate perceived grounding, with participants reporting fewer human review iterations. We outline operational lessons and limitations, emphasizing reproducibility, auditability, and pathways to broader standards and assisted patching.
\end{abstract}
]



\printAffiliationsAndNotice{} 

\section{Introduction}

Modern software organizations increasingly rely on pull request (PR) centric workflows to maintain development velocity while protecting code quality. Yet code reviews, which are a core requirement in software development, remain a human-intensive and error-prone process that consumes a significant portion of engineering time \cite{bosu2015characteristics, thomson2021static}. Traditional static analyzers integrated into PRs, such as Google's Tricorder or integrations available through tools like CodeQL \cite{youn2023declarative} and Semgrep \cite{bennett2024semgrep}, are crucial for enforcement. However, they often surface large volumes of findings with limited explanatory rationale, creating code triage burdens and inadvertently contributing to developer warning fatigue \cite{sadowski2015tricorder, youn2023declarative, NEURIPS2023_662b1774}. Conversely, while large language models (LLMs) excel at explaining code, identifying bugs, and suggesting remedies, their naive application in continuous integration and continuous delivery (CI/CD) pipelines remains a challenge. Without a verifiable chain of evidence, they are susceptible to misidentification, hallucinations, inconsistent reasoning, and prohibitive operational costs, which fail to deliver the concise, trustworthy, and timely feedback required for enterprise adoption \cite{zhang2025llm}.

This paper presents a production system for grounded AI code review that couples the determinism of industry-grade static analysis with the explanatory power of LLMs, all supported by a resource-efficient model serving architecture. The system is designed to deliver actionable, PR-native feedback at a medium enterprise scale, covering dozens of repositories and hundreds of weekly pull requests. Our central design principle is grounding: every LLM-generated explanation is explicitly anchored to concrete, verifiable evidence. This includes compiler-verified builds, specific static-analysis findings, formal rule definitions, and precise file/line locations. This approach forces the model to reason about the code in the context of a tangible compliance or quality requirement, rather than reasoning about the code in general. To keep prompts compact yet semantically rich, the system performs abstract syntax tree (AST) and call-graph guided context extraction, selecting only the functions, types, and lines necessary to understand a finding. We implement it as a "grounding first, then generate" pattern, which has shown success in recent repository-context and graph-guided approaches. This approach effectively constrains the model's reasoning space to reduce hallucinations and enhance the relevance of its fix suggestions \cite{xie2025core}.

We evaluate our approach on internal codebases representative of medium-scale enterprise development using an offline, reproducible benchmark at the hunk level. Our contributions are fourfold: \textbf{(i) Hybrid Grounding Methodology} pairs static-analysis evidence with LLM explanations to produce citation-rich PR comments; \textbf{(ii) Single-GPU Resource-Efficient Serving} delivers a p50 first-feedback proxy of 59.8s (Table~\ref{tab:profile_summary}) on a quantized open-weight model; \textbf{(iii) Enterprise Integration Blueprint} captures build orchestration, deviation policy handling, audit provenance, and reproducible prompts; and \textbf{(iv) Competitive On-Prem Effectiveness} shows a 6-bit Qwen2.5 coder profile matching larger APIs while lowering rule violation introductions (Figure~\ref{fig:benchmark_grid}, Table~\ref{tab:profile_summary}) and preserving on-prem control. Section~\ref{sec:methodology} expands on the architecture and serving pipeline.

The remainder of this paper details the system architecture and methods, presents quantitative and qualitative results, situates the work within the context of related research and tools \cite{rasheed2024ai}, and outlines limitations and future directions, including automated patching, broader standards support, and IDE-time assistance.

\section{Background}

\subsection{Large Language Models for Developer Assistance}
LLMs have emerged as a powerful tool for developer assistance, capable of explaining complex code, summarizing changes, and proposing concrete fixes. A new generation of PR-native tools, including GitHub Copilot for Pull Requests and CodeRabbit, leverages this capability to provide automated, natural-language feedback directly within the review process \cite{goccmen2025enhanced, cihan2025automated}. These tools demonstrate the potential of AI to reduce the cognitive load on human reviewers, especially for issues related to style, clarity, and simple defects.

However, the unconstrained application of LLMs in the high-stakes environment of code review presents significant risks. Without guardrails, these models are prone to hallucinations, where they confidently invent incorrect facts or identify non-existent issues \cite{zhang2025llm}. They might provide broad, out-of-context suggestions that don't apply, weakening trust with developers. Furthermore, the operational realities of enterprise CI/CD, including strict latency service level agreements (SLAs) and service level objectives (SLOs), spiky traffic patterns, and constrained budgets, make the naive deployment of large models computationally infeasible.

\subsection{Grounding: A Hybrid Approach to Trustworthy AI Review}
The most effective path to mitigating these risks is grounding: constraining an LLM's reasoning to a foundation of verifiable evidence. In the context of code review, this means every AI-generated comment must be explicitly anchored to deterministic signals, such as compiler-verified builds, static-analysis findings, formal rule definitions, and precise file/line locations. This "grounding-first" approach forces the model to reason about a specific, verifiable issue rather than speculate about the code in general.

This hybrid model pairs analyzer recall and localization with LLM contextual explanation. For each finding, we extract a minimal code segment along with the rule rationale, forming a compact prompt that preserves necessary semantics while minimizing tokens. The result is fewer, higher-quality comments that explain why an issue matters and how to remediate it.

\subsection{Resource-Efficient LLM Serving for CI/CD Workloads}
Delivering grounded feedback within the demanding SLOs of a CI/CD pipeline requires a purpose-built serving architecture. Enterprise environments often rely on a limited pool of shared GPUs, necessitating a focus on resource efficiency. To be practical, a serving stack must handle the bursty, diurnal traffic patterns of development activity, ensuring a predictable and low "time-to-first-comment" (e.g., near sub-minute) \cite{walkowiak2024assessing, lou2025towards}.

Key techniques include quantization (reducing footprint for single-GPU fit), KV/prefix and response caching to avoid repeat work, and an on-demand lifecycle that unloads during idle periods. The implementation on AMD Instinct + ROCm is hardware-agnostic, avoiding vendor lock-in.

\paragraph{Requirements for an Enterprise-Grade System}

Based on this review of the state-of-the-art (SOTA) and its limitations, we derive four core design requirements for a practical, PR-native AI code review system: \textbf{R1 — Grounded Accuracy} anchors comments in verifiable evidence (AST spans, rule citations, tool findings) with precise line references; \textbf{R2 — CI-Grade Latency and Throughput} targets sub-minute (P95) feedback for typical PRs, excluding analyzer build time, even under bursty CI/CD loads; \textbf{R3 — Cost Efficiency} keeps the stack on a single shared enterprise GPU via aggressive caching and on-demand allocation; and \textbf{R4 — Enterprise Operability} delivers PR-native integration, deviation policy support, and the observability, security, and audit trails required in enterprise environments \cite{liargkovas2023quieting}.

We evaluate safety-oriented C/C++ standards (e.g., MISRA C/C++) as a case study, but the architecture is standard-agnostic and applies broadly to organizational reviews requiring robust, low-resource serving, and verifiable outputs \cite{vero2025baxbench}.

\section{Methodology}
\label{sec:methodology}

This section details the architecture and operational flow of our grounded AI code review system. We first provide a high-level overview, followed by a detailed examination of the core components responsible for orchestration, context extraction, prompt engineering, and resource-efficient model serving.

Enterprise deployment imposes hard constraints that typical LLM demonstrations ignore: strict end-to-end latency targets, volatile and spiky traffic patterns aligned with business hours, limited and shared graphical processing unit (GPU) resources, and stringent data-residency and security requirements. We address these constraints with a serving stack purpose-built for the CI/CD environment. The architecture features a decoupled message queue to buffer and route incoming requests, enabling graceful handling of traffic bursts. This queue feeds a multi-tier caching system and a dynamic router that forwards only novel, uncomputed requests to a single GPU inference service. The service runs a quantized large model (e.g., 4--8 bits using llama.cpp) with on-demand loading to fit on a single enterprise GPU and reclaim compute hours when pipelines are quiet. By leveraging both prompt-level KV/prefix and response-level caching, the system minimizes re-computation. It ensures that the GPU is reserved for generating new insights, dramatically improving throughput and end-to-end latency.

We instantiate this architecture in a GitHub-integrated bot ("Auto CodeReview") that (1) orchestrates repository-specific builds and static analysis checks, (2) extracts minimal, high-signal context around each finding, (3) synthesizes structured prompts that include rule rationales and protective guardrails, and (4) posts structured PR comments that explain why an issue matters and how to fix it, complete with links back to the source code and policy documents. The system emits both violation reports for auditability and PR-native comments for seamless workflow integration. It is designed to operate alongside existing PR assistants (e.g., Copilot for Pull Requests, CodeRabbit) while specializing in high-trust, rule-aware feedback. While our primary case study targets a safety-oriented C/C++ rule set (specifically motor industry software reliability association (MISRA) \cite{hatton2004safer}), the architecture is standard-agnostic and can be readily applied to organization-specific policies and security patterns across any language.

\subsection{System Overview}

\begin{figure*}[h]
 \centering
 \includegraphics[width=0.9\textwidth]{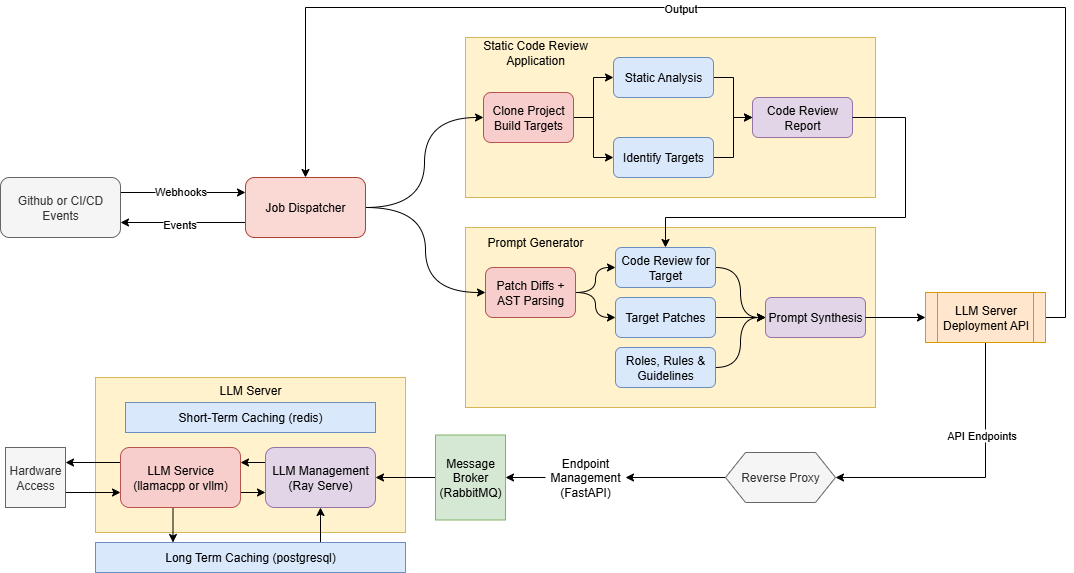}
 \caption{Complete end-to-end framework of AutoCodeReview: The system consists of Code-Review Orchestrator, which extracts, analyzes, and generates code review prompts (Static Analyzer + Prompt Generator), and LLM Serving Backend, which provides access to LLM Service.}
 \label{fig:autocodereview}
\end{figure*}

The system is architected as two cooperating microservices designed for modularity and scalability within an enterprise CI/CD environment, shown in Figure \ref{fig:autocodereview}. 
\begin{enumerate}
 \item \textbf{Code-Review Orchestrator:} A GitHub App, implemented in Node.js and managed by the PM2 process manager, serves as the primary control plane. It listens for PR webhooks, manages the state of review jobs, and orchestrates the entire workflow. Its responsibilities include checking out the relevant repository state, executing repository-specific build and static analysis commands, parsing the resulting reports, performing context extraction, synthesizing prompts for the LLM, and posting structured, threaded comments back to the GitHub PR.
 \item \textbf{LLM Serving Backend:} A containerized, on-premise service built with FastAPI and Ray Serve provides the generative capabilities. It is designed for high-throughput, low-latency inference under the bursty traffic patterns of CI/CD. The backend hosts a quantized LLM, managed by llama.cpp and Ray Serve for on-demand loading and unloading. To maximize resource efficiency, it employs a multi-tier caching architecture (in-memory KV/prefix, Redis, and PostgreSQL) and a RabbitMQ message queue for intelligent request routing and load leveling. An Nginx reverse proxy manages TLS termination and secure ingress.
\end{enumerate}

\subsubsection{End-to-End Workflow}

The system operates as a stateful pipeline triggered by PR events. The end-to-end flow proceeds as follows: (1) A developer opens or updates a PR, triggering a GitHub webhook received by the orchestrator. (2) The orchestrator checks out the specified commit and, using a repository-specific configuration, executes the necessary build commands. (3) The configured static analyzer is invoked on the built code, generating a structured report of potential violations. (4) The orchestrator parses this report, filtering for findings relevant to the PR's diff. (5) For each relevant finding, an AST-guided context extractor selects a minimal, high-signal code snippet and enriches it with rule information. (6) A structured, grounded prompt is assembled and sent to the LLM serving backend. (7) The backend first checks its multi-tier cache; on a cache miss, the request is routed to the LLM for generation. (8) The generated explanation, including rationale, risk, and remediation advice, is returned to the orchestrator, which formats and posts it as a threaded PR review comment. (9) On subsequent commits, the system cleans up stale comments and archives the interaction.

\subsection{Orchestrator: Builds, Analysis, and Policy}

\textbf{Config-driven Builds:} To support a heterogeneous multi-repository environment, each project defines its build and analysis configuration in a repository-local JSON file. This file specifies the toolchain (e.g., GCC, specific cross-compilers), build flags, and the static analysis standard to apply (e.g., MISRA C:2012). The orchestrator uses this configuration to create a clean, reproducible build environment and invoke the static analyzer with the correct rule packs enabled. Where supported, the system defaults to a diff-scoped or incremental analysis to minimize latency.

\textbf{Rule Registry and Enrichment:} Our implementation leverages the Coverity static analyzer. The orchestrator invokes cov-build and cov-analyze to generate an intermediate data format. From this, a machine-readable JSON report is produced. This report is parsed to extract detailed information for each finding, including the rule ID (e.g., MISRA Rule 10.1), violation description, severity, and Coverity's suggested remediation. The orchestrator then filters these findings, retaining only those whose file and line locations fall within the PR's changed code patch. This ensures that every comment is directly relevant to the author's modifications.

\textbf{Deviation Management:} To handle necessary exceptions to static analysis rules, the system supports a formal deviation process. While Coverity has internal mechanisms for managing deviations, our system uses a supplementary JSON-based policy file maintained within each repository. This file records approved exceptions, specifying the scope (file, function, and rule), a rationale for the deviation, and the owner. When the orchestrator processes an analysis report, it cross-references findings against this policy file. Matching findings are suppressed from the PR comments but are preserved in the full audit logs to maintain compliance traceability.

\subsubsection{Token-Budgeted Context Extraction}
To create prompts that are both compact and semantically rich, the system employs a multi-faceted context extraction strategy that operates within a fixed token budget. For each finding identified in the static analysis report, the extractor uses the diff to locate the change, parses the AST to understand structure, and gathers the enclosing function body (with non-essential comments collapsed), the direct caller and callee names that drive the relevant flow, and if space remains tight a sliding window of $\pm k$ lines around the violation. The final extracted snippet is annotated, giving the LLM a complete and verifiable view of the code in question. A high-level description of the review pipeline is shown in algorithm \ref{alg:autocodereview}.

\begin{algorithm}[htbp!]
\caption{High-Level Algorithm for a Scalable, Asynchronous Auto Code Review System.}
\label{alg:autocodereview}

\Proc{AutoCodeReview{$W$}}{
 $P \gets \ExtractPayload(W)$\;
 $T \gets \CreateTask(P)$\;
 \Enqueue{$Q, T$} \tcp*{asynchronous processing}
}

\BlankLine

\Proc{ProcessTask{$T$}}{
 $R_{info} \gets T.\mathrm{repo\_info}$\;
 $C_{local} \gets \CloneOrUpdateRepo(R_{info}, T.\mathrm{pr\_number})$\;
 $B_{config} \gets \GetBuildConfig(R_{info})$\;
 $(R_{SA}, G_{call}) \gets \RunStaticAnalysis(C_{local}, B_{config})$ \tcp*{e.g., Coverity, Doxygen}
 $D_{chunks} \gets \GetDiffChunks(T.\mathrm{diff})$\;
 $V_{grouped} \gets \GroupViolationsByChunk(R_{SA}, D_{chunks})$\;

 \ForEach{$v \in V_{grouped}$}{
 $C_{snip} \gets \ExtractCodeSnippet(C_{local}, v.\mathrm{location})$\;
 $C_{ctxt} \gets \GetCodeContext(C_{snip}, G_{call})$ \tcp*{sliding window, AST, call hierarchy}
 $\mathbb{P}_{LLM} \gets \GenPrompt(\mathrm{role, rules}, v, C_{snip}, C_{ctxt})$\;
 $S_{LLM} \gets \QueryLLM(\mathbb{P}_{LLM})$ \tcp*{on-prem LLM endpoint}
 \AppendToReport{$S_{LLM}$}\;
 }

 $C_{PR} \gets \FormatPRComment(\mathrm{AggregatedReport})$\;
 \PostCommentToPR{$R_{info}, T.\mathrm{pr\_number}, C_{PR}$}\;
 \Cleanup{$C_{local}$} \tcp*{remove local clone}
}
\end{algorithm}

\subsubsection{Prompting and Guardrails}

All prompts adhere to a structured schema designed to enforce grounding and elicit high-quality, actionable responses. Key components include \textbf{Role \& scope}, which instructs the LLM to behave as a senior compliance reviewer focused on supplied context; \textbf{Rule Rationale}, a concise rule explanation pulled from analyzer documentation; \textbf{Finding Metadata}, covering rule ID, file path, line numbers, and the raw analyzer message; the annotated \textbf{Code Snippet} itself; an \textbf{Output Contract} that demands rationale, risk framing, and remediation options with explicit line citations; and explicit \textbf{Guardrails} that forbid speculation beyond the provided snippet. Responses are programmatically checked against this schema, then grouped by rule and posted as threaded PR review comments for readability.

\subsection{LLM Serving: Efficient Operation in CI/CD}

 \textbf{Serving Stack:} The LLM serving backend is a Dockerized on-premise deployment running on an AMD MI210 GPU with the ROCm stack \cite{zhang2024llmcompass}. It consists of four integrated services: (1) an Nginx reverse proxy for TLS termination and ingress control; (2) PostgreSQL for persistent, long-term response caching and analytics; (3) RabbitMQ for message broking, which decouples the API from the inference workers and buffers requests during traffic spikes; and (4) the LLM Server itself, which uses FastAPI for the API layer, Redis for a low-latency cache, and Ray Serve to manage the lifecycle and deployment of the llama.cpp inference engine (Algorithm \ref{alg:llm_server_short}).

 \textbf{Model and Quantization:} We deploy a code-specialized 32-billion parameter model (a fine-tuned variant of Qwen2.5) in a 6-bit GGUF quantized format. Quantization reduces the VRAM footprint from approximately 64 GB for a bfloat16 model to under 24 GB, allowing it to fit comfortably on a single enterprise GPU with minimal-to-no perceptible degradation in quality for this explanatory task.

 \textbf{On-demand Lifecycle:} To maximize resource utilization in a shared compute environment, the model is loaded into VRAM only upon the arrival of the first request in an empty queue (a cold start of $\sim 30$--45 seconds) \cite{fu2024serverlessllm}. Ray Serve automatically unloads the model from the GPU after a configurable idle timeout (e.g., 30 minutes), returning the GPU capacity to the shared pool. Subsequent requests arriving while the model is warm bypass this loading cost entirely.

\begin{algorithm}[htbp!]
\caption{Multi-Tier Cached LLM Serving. Symbols: $R$: Request, $P$: Payload, $C_1$: L1 Cache (In-Memory), $C_2$: L2 Cache (Persistent), $MQ$: Message Queue, $k_1$: L1 Key, $j_{id}$: Job ID, $M$: LLM Model, $L_{gpu}$: GPU Lock.}
\label{alg:llm_server_short}

\Require{An HTTP Request $R$.}
\Ensure{An HTTP Response.}

\Proc{HandleRequest{$R$}}{
 Authenticate $R$; \If{fails}{\textbf{reject}\;}
 $k_1 \gets \GenerateKey(R)$\;
 $Resp \gets C_1.\mathrm{get}(k_1)$\;
 \If{$Resp \neq \mathrm{NULL}$}{\Return{$Resp$} \tcp*{L1 Cache Hit}}
 \tcp{L1 Miss: delegate to message broker}
 $P \gets \CreatePayload(R)$\;
 $(Q_r, c_{id}) \gets MQ.\mathrm{setup\_callback}()$\;
 $MQ.\mathrm{publish}(P, Q_r, c_{id})$\;
 $Resp \gets$ \textbf{await }$MQ.\mathrm{get\_response}(Q_r, c_{id})$\;
 $C_1.\mathrm{set}(k_1, Resp)$\;
 \Return{$Resp$}\;
}

\BlankLine

\Proc{ProcessTask{$P$}}{
 $T_{in} \gets \GeneratePrompt(P)$\;
 $j_{id} \gets \mathrm{hash}(T_{in})$\;
 $Res_c \gets C_2.\mathrm{get}(j_{id})$ \tcp*{Check L2 Cache}
 \If{$Res_c = \mathrm{COMPLETE}$}{
 $Res \gets \Format(Res_c)$ \tcp*{L2 Cache Hit}
 }\Else{
 $C_2.\mathrm{create\_pending}(j_{id}, P)$\;
 $Res \gets$ \ProcessLLM{$T_{in}, j_{id}$}\; 
 }
 $MQ.\mathrm{publish\_response}(P.\mathrm{reply\_to}, Res, P.c_{id})$\;
}

\BlankLine

\Fn{ProcessLLM{$T_{in}, j_{id}$}}{
 Acquire $L_{gpu}$\;
 \If{$M$ is UNLOADED}{$M.\mathrm{load}()$ \tcp*{Cold Start}}
 \try{}{
 $O_{llm} \gets M.\mathrm{inference}(T_{in})$\;
 $C_2.\mathrm{update}(j_{id}, \texttt{'COMPLETE'}, O_{llm})$\;
 $Res \gets \Format(O_{llm})$\;
 }
 \catch{Exception $e$}{
 $C_2.\mathrm{update}(j_{id}, \texttt{'ERROR'}, e)$\;
 $Res \gets \FormatError(e)$\;
 }
 \finally{}{
 Release $L_{gpu}$\;
 $M.\mathrm{reset\_unload\_timer}()$ \tcp*{Keep model warm}
 }
 \Return{$Res$}\;
}
\end{algorithm}

\textbf{Throughput and Caching:} The system uses multiple strategies to maximize throughput. Ray Serve minimally batches incoming requests to improve GPU utilization. A multi-tier cache minimizes redundant computation: a KV/prefix cache in llama.cpp accelerates prompts with shared prefixes (common when analyzing multiple violations of the same rule) \cite{kwon2023efficient}, a Redis instance caches complete responses for several hours, and the PostgreSQL database provides a long-term, persistent cache of hashed request-response pairs, deduplicating common violations across the organization over weeks or months.

\textbf{Reliability:} All inference requests have a client-side timeout (e.g., 300 seconds) and employ an exponential backoff retry strategy to handle transient GPU or network failures. The RabbitMQ queue provides backpressure signals to the orchestrator during sustained high load. The system is designed to fail closed: in the event of a timeout or unrecoverable error, it logs the failure and posts no comment, rather than providing partial or ungrounded feedback.

\subsubsection{Observability, Security, and Audit}
 \textbf{Tracing and Metrics:} A lightweight Streamlit dashboard provides real-time observability by querying the PostgreSQL database. It visualizes key performance indicators across the entire lifecycle, including latency breakdowns, GPU utilization, error/timeout rates, and relevant metrics for deployment.

 \textbf{Security:} The serving API is isolated on a private network, accessible only by the orchestrator via an API and basic authentication. All source code, analysis reports, and prompts remain on-premise, satisfying strict enterprise data residency and security requirements.

 \textbf{Auditability:} For compliance and traceability, every PR comment includes a detailed provenance record: the full commit SHA, file path, line range, rule ID, and the version of the model used for generation. All deviation decisions and comment lifecycles are persistently logged for periodic compliance reviews.

\subsection{Dataset and Benchmark Construction}
\label{sec:dataset}
Our evaluation relies on a reproducible, hunk-level benchmark derived from 10 medium-scale C/C++ repositories. Seven are openly available upstream (diverse domains: embedded utilities, cryptography/math, protocol tooling, and compression/hash libraries), two are internal variants (forked adaptation), and one is an entirely internal component selected to reflect typical enterprise maintenance patterns (a mix of legacy and actively evolving subsystems). Combined, the snapshot corpus spans roughly 600K lines of code after preprocessing (comments preserved for context extraction; generated/build artifacts excluded). 

To approximate a realistic PR structure while enabling deterministic replays, we synthesize 100 PR scenarios that expand to 314 atomic hunks (a contiguous diff region plus its associated static-analysis findings). Hunks preserve: (a) original file paths and line spans, (b) rule identifiers and severities emitted by the analyzer, and (c) sufficient unchanged surrounding code for AST-guided context extraction. This ensures prompts constructed offline mirror production grounding behavior (same token-shaping heuristics, rationale embedding, and rule metadata). 

\paragraph{Generation Pipeline.} An idempotent pipeline produces the benchmark: (1) clone or refresh each repository at a pinned commit; (2) enumerate candidate modification sites by sampling existing static-analysis findings and representative clean regions; (3) apply controlled source edits (e.g., introduce or remediate patterns, adjust control flow, perturb literals) to create prospective pre/post states; (4) rebuild and re-run static analysis to capture rule deltas; (5) segment resulting diffs into atomic hunks; (6) serialize per-hunk metadata (pre/post violation vectors, file context slices, rule rationale, diff stats) into data artifacts consumed by the evaluation harness. Each step records inputs (commit SHA, tool version, transformation seed), enabling exact regeneration. 

\paragraph{Open vs Internal Balance.} Internal repositories contribute enterprise-specific idioms (custom build flags, legacy macro usage) that stress the robustness of context extraction; open-source projects supply heterogeneity and external reproducibility. We deliberately cap internal proportion to avoid overfitting architectural claims to proprietary structure while still exercising on-prem build constraints. 

\subsection{Experimental Design and Metrics}

We evaluate and compare the system on 10 C/C++ codebases using a purpose-built offline benchmark that we constructed specifically to exercise the end-to-end grounded review pipeline in a controlled, repeatable setting. The benchmark consists of 100 synthetic-but-replayable pull request submissions expanded into a total of 314 atomic hunks (each hunk = a contiguous diff segment with associated static-analysis findings), covering roughly 600K lines of code. Each hunk is processed by an automated harness that mirrors the production orchestrator's core phases: isolated workspace materialization, context extraction, structured prompt generation, LLM invocation, patch application, rebuild + static re-analysis, and rule delta computation. All runs are performed on a single-GPU node (AMD MI210 with 64 GB HBM, 64 vCPUs, 512 GB RAM) to provide a stable hardware baseline.

This "AutoCodeReview" hunks benchmark is intentionally PR-shaped: hunks preserve original file paths, line spans, and associated rule metadata so that prompt construction and grounding logic behave as they would inside a live CI/CD pull request workflow. However, two important distinctions from real-time usage apply: (i) the offline harness automatically applies candidate patches and evaluates post-build rule deltas, whereas the deployed system surfaces suggestions (non-destructive) for human acceptance; and (ii) interactive ergonomic optimizations (deferred comment posting, UI pacing, incremental diff scoping, developer-driven retry) are not modeled. Consequently, the latency and violation-reduction numbers reported here may differ slightly from those in production, and acceptance-oriented metrics (such as developer adoption) are not directly observable in this setting.

Metric aggregation is performed in two passes: a first pass that collects raw per-hunk outcomes (status, rule deltas, diff characteristics, timing breakdown), and a second pass that derives higher-level reporting metrics (effective fix rate, coverage/introduced fractions, latency percentiles including first-feedback, edit efficiency, ablation deltas, competitive outcomes). Unless otherwise noted, all metrics refer to the automated benchmark environment (upper-bound capability; not interactive wall-clock guarantees).

Our evaluation focuses on computable metrics generated by the harness: violation reduction, rule coverage, introduced violations, effectiveness proxies, competitive outcomes, latency (including first-feedback), edit efficiency, prompt compactness, ablation deltas, and directional survey perception signals.

\section{Results}

\begin{figure*}[ht]
 \centering
 \begin{subfigure}[t]{0.45\textwidth}
  \centering
  \includegraphics[width=\linewidth]{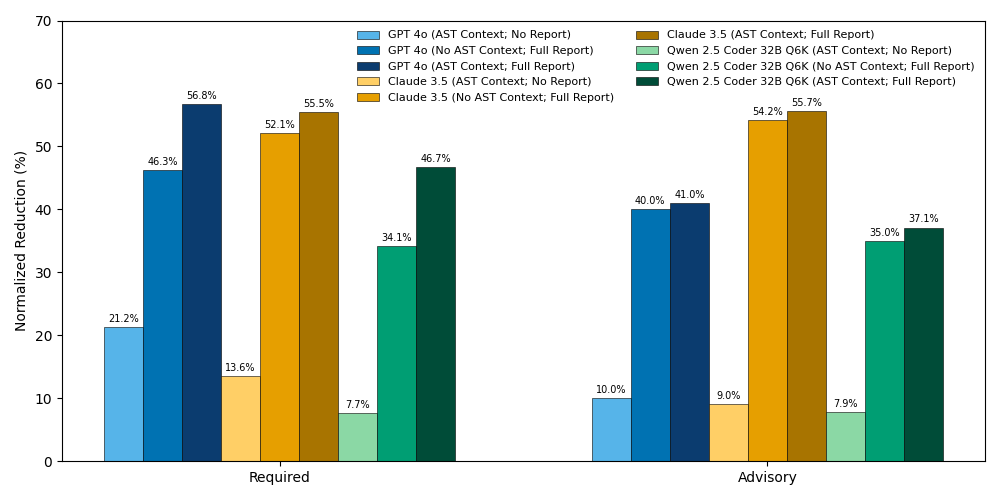}
  \caption{Normalized reductions by severity.}
  \label{fig:severity_reduct}
 \end{subfigure}\hfill
 \begin{subfigure}[t]{0.45\textwidth}
  \centering
  \includegraphics[width=\linewidth]{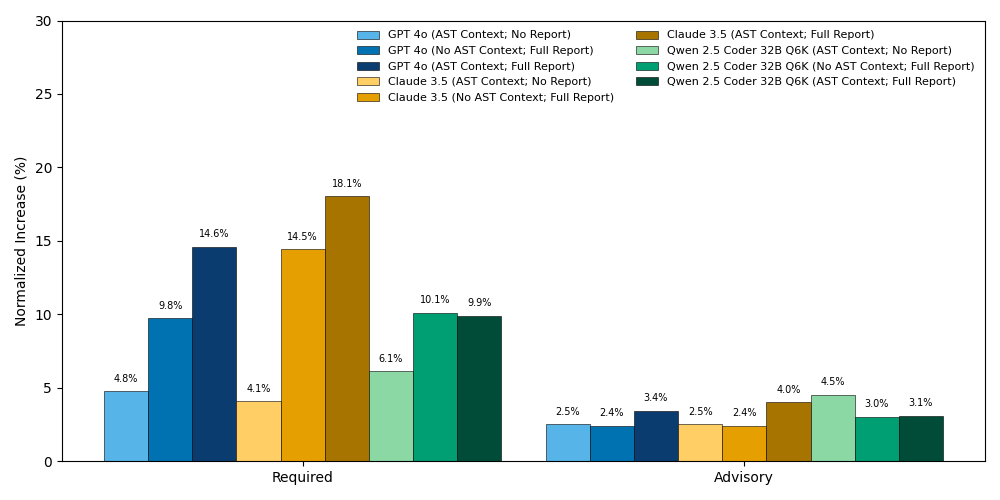}
  \caption{Normalized increases/introductions.}
  \label{fig:severity_increase}
 \end{subfigure}
 \begin{subfigure}[t]{0.48\textwidth}
  \centering
  \includegraphics[width=\linewidth]{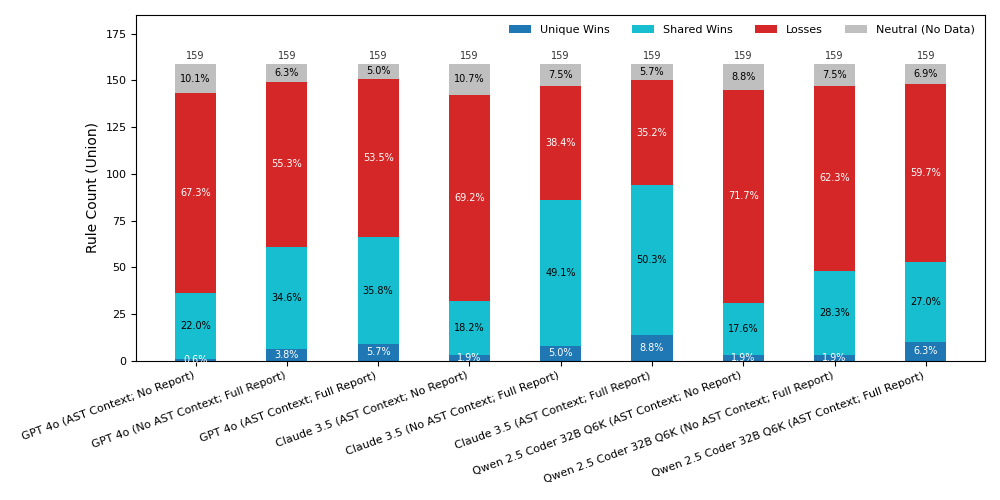}
  \caption{Union-normalized per-rule outcomes.}
  \label{fig:rule_outcomes}
 \end{subfigure}\hfill
 \begin{subfigure}[t]{0.48\textwidth}
  \centering
  \includegraphics[width=\linewidth]{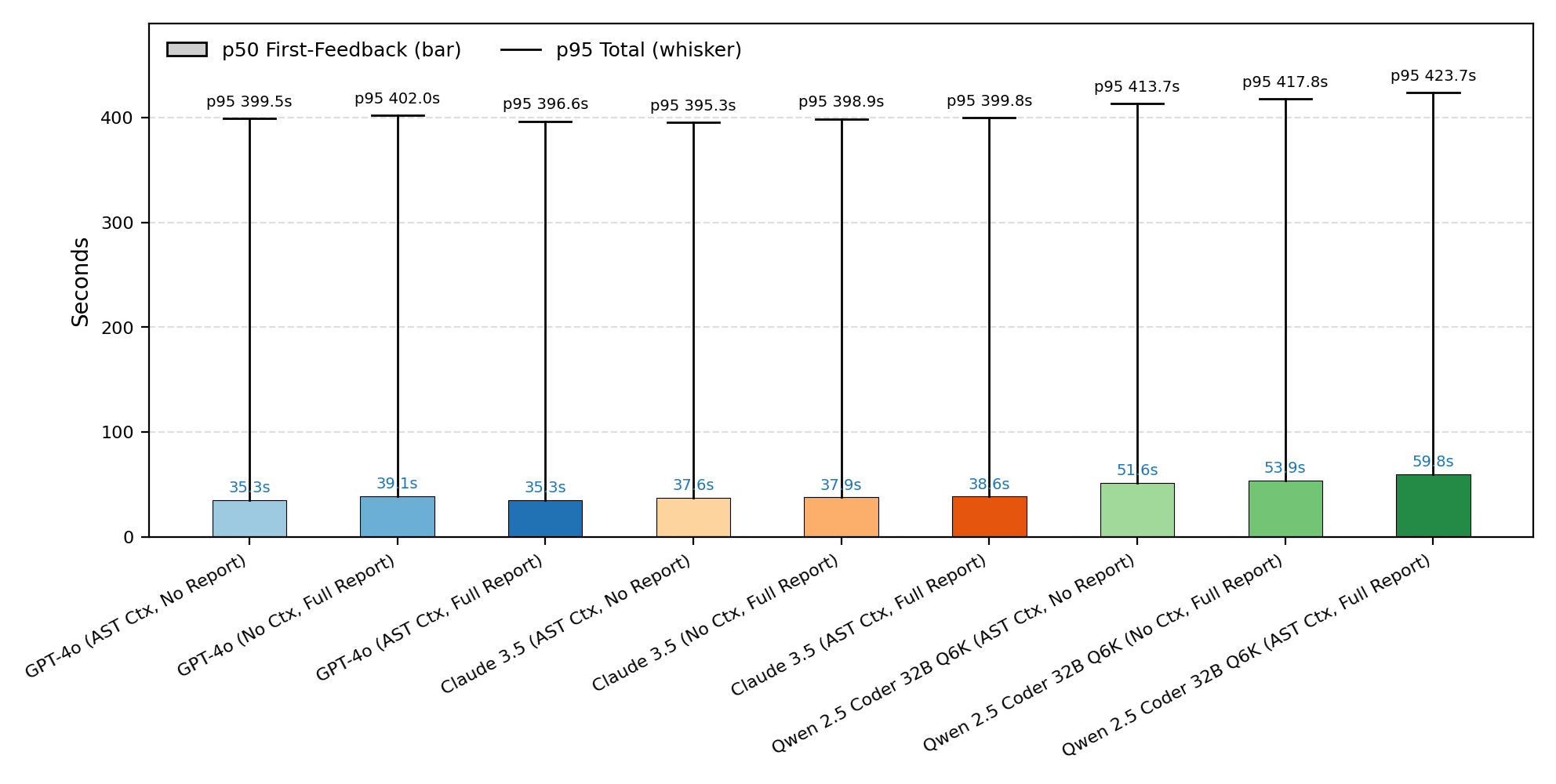}
  \caption{Latency (p50 total, p95 whiskers, first-feedback).}
  \label{fig:latency_breakdown}
 \end{subfigure}
 \caption{Benchmark summary: severity-level reductions and introductions (top), per-rule outcome decomposition (bottom-left), and latency characteristics (bottom-right). Reduction = (pre-post)/pre over violations with pre>0; new-only violations (pre=0, post>0) contribute 1.0 to introductions.}
 \label{fig:benchmark_grid}
\end{figure*}

\begin{table*}[htbp!]
 \centering
 \caption{Per-profile summary of key metrics: reduction ratios, coverage, recall proxies, latency (p50 total and first-feedback), and edit efficiency.}
 \label{tab:profile_summary}
 \resizebox{\textwidth}{!}{\begin{tabular}{llrrrrrrrrr}
  \toprule
 Model & Setting & Reduction. & NetRed & CovFrac & IntroFrac & WtdRPx & MacRPx & p50Tot(s) & p50FF(s) & Lines/Vio \\
  \midrule
 Claude-3.5 & Ctx,NoRpt & 0.127 & 0.066 & 0.632 & 0.500 & 0.145 & 0.233 & 37.64 & 37.64 & 2.59 \\
 Claude-3.5 & NoCtx,Full & 0.471 & \textbf{0.305} & \textbf{0.904} & 0.456 & 0.593 & \textbf{0.644} & 37.92 & 37.92 & \textbf{2.28} \\
 Claude-3.5 & Ctx,Full & \textbf{0.482} & 0.290 & 0.897 & \textbf{0.471} & \textbf{0.617} & 0.635 & 38.62 & 38.62 & 2.29 \\
 Gpt-4o   & Ctx,NoRpt & 0.183 & 0.112 & 0.662 & 0.515 & 0.203 & 0.287 & \textbf{35.26} & \textbf{35.26} & 5.14 \\
 Gpt-4o   & NoCtx,Full & 0.408 & 0.284 & 0.824 & 0.537 & 0.493 & 0.505 & 39.06 & 39.06 & 3.21 \\
 Gpt-4o   & Ctx,Full & 0.456 & 0.285 & 0.882 & 0.603 & 0.576 & 0.572 & 35.30 & 35.30 & 3.02 \\
 Qwen2.5-coder-23b  & Ctx,NoRpt & 0.095 & 0.005 & 0.574 & 0.581 & 0.112 & 0.201 & 51.63 & 51.63 & 5.76 \\
 Qwen2.5-coder-23b  & NoCtx,Full & 0.332 & 0.192 & 0.794 & 0.574 & 0.409 & 0.448 & 53.88 & 53.88 & 3.34 \\
 Qwen2.5-coder-23b  & Ctx,Full & 0.410 & 0.276 & 0.772 & 0.596 & 0.493 & 0.446 & 59.81 & 59.79 & 2.88 \\

  \bottomrule
 \end{tabular}}
 \vspace{0.3em}\footnotesize
 Reduction.: overall violation reduction ratio = $(\text{pre}-\text{post})/\text{pre}$ aggregated over violations with pre>0 (introduced-only rules excluded from the numerator); NetRed: net reduction after subtracting introduced violations (accounts for regressions); CovFrac: fraction of distinct pre-existing rules for which at least one violation was reduced; IntroFrac: fraction of distinct rules that were newly introduced or increased; WtdRPx: weighted recall proxy = (sum of per-rule reductions in counts)/(sum of pre counts) (upper-bound recall; precision unmeasured); MacRPx: macro (unweighted) recall proxy = mean over rules with pre>0 of per-rule reduction ratios; p50Tot: median (p50) end-to-end pipeline time per hunk (s); p50FF: median first-feedback proxy (build + LLM phases only) (s); Lines/Vio: average changed lines per violation removed; Setting: context (Ctx vs NoCtx) and report formatting (Full vs NoRpt) configuration. All proxies omit precision because dismissals / false positives are not observable offline.
\end{table*}

This section reports benchmark outcomes (automated; upper-bound capability) plus directional survey perceptions. All quantitative metrics derive from evaluation harness; survey data is qualitative only.

\subsection{Severity-Level Normalized Violation Reductions and Regressions}
The top row of Figure~\ref{fig:benchmark_grid} reports two severity-level metrics: normalized violation reductions (left) and normalized regressions (right). Let \textit{pre} and \textit{post} denote the number of violations per item at a given severity level before and after the LLM-based fix, respectively. For items with \textit{pre} > 0, reduction is computed as $(\text{pre}-\text{post})/\text{pre}$; positive values indicate improvement, negative values indicate regressions (i.e., increases where $\text{post} > \text{pre}$). For items with \textit{pre} = 0 and \textit{post} > 0, we record an introduction of 1.0 in the right panel. Items with \textit{pre} = \textit{post} = 0 are excluded to avoid inflating apparent effectiveness with trivial no-change cases. This separation makes regressions explicit while preventing inflation from absent baselines.

\subsection{Union-Normalized Per-Rule Outcomes}
The bottom-left panel of Figure~\ref{fig:benchmark_grid} reports per-rule win rates across models/scenarios under a union-normalized evaluation. For each rule in the union of all rules touched by any model, we compare outcomes pairwise and compute the fraction of rules where a given model outperforms its comparator. Rules untouched by a model in the post-fix are treated as neutral (no win/loss), ensuring that unique and shared coverage are comparable and that non-participation is distinguished from underperformance.

\subsection{Latency, First-Feedback, and Throughput}
The bottom-right panel of Figure~\ref{fig:benchmark_grid} and Table~\ref{tab:profile_summary} summarize timing and throughput. We decompose end-to-end wall-clock latency into four phases: context preparation, LLM inference, apply (patch/format), and build/analysis; total latency is the sum of these phases. First-feedback latency is defined as LLM + build/analysis at the atomic-hunk level, the earliest point at which an actionable comment can be emitted. At the PR level, p95 total denotes the 95th percentile of end-to-end completion time across PRs. Throughput is reported as items per hour (PRs/hour and hunks/hour), measured under the benchmark’s concurrency settings. Table~\ref{tab:profile_summary} provides per-phase statistics, PR-level totals (median and p95), and the corresponding throughput figures.

\subsection{Ablations and Context Economy}
Context and report-format ablations showed that trimming structured context reduced token footprint but degraded recall proxies more than it improved latency, supporting retention of current context heuristics.

\subsection{Developer Workflow Impact (Survey, Directional)}
The internal survey (n=8) indicated directional (non-inferential) signals about workflow impact. Respondents reported: mean self-reported time-to-first-feedback of 2.75 minutes (P75 = 4.0), immediate adoption of $\sim 50\%$ of surfaced suggestions, and an overall adapt-or-accept rate of $\sim 56\%$ after iterative refinement. The median perceived clarity was 4/5, and grounding was rated 3.38/5. 57\% of participants indicated fewer human review iterations after adoption (none reported an increase). Qualitative free-text responses cited a lower cognitive load compared to raw static-analyzer output (less time spent triaging warnings) and expressed a desire for optional, safety-gated patch suggestions. These perception metrics are directional only and motivate future objective instrumentation; they are not used to instantiate statistical claims.

\section{Discussion}

The evaluation demonstrates that a grounded hybrid design can provide rule-aware, low-latency review assistance on a single shared GPU, while narrowing the performance gap with larger proprietary APIs. We now discuss the results in terms of design goals, generality, adoption factors, and remaining risks.

\subsection{Grounding Drives Actionable Precision}
Anchoring every comment to a concrete static-analysis finding (rule ID + file/line) focuses the LLM on targeted rationale rather than unconstrained defect hunting, reducing hallucination and triage overhead. In Table~\ref{tab:profile_summary}, grounded configurations consistently improve recall proxies (e.g., GPT-4o WtdRPx: 0.493$\rightarrow$0.576; Qwen2.5 WtdRPx: 0.409$\rightarrow$0.493) and net reduction (Qwen2.5 NetRed: 0.192$\rightarrow$0.276), with modest latency trade-offs. AST-guided, token-budgeted context provides sufficient semantics without prompt bloat; ablations that remove structured context degraded recall proxies more than they improved latency, supporting the current heuristics.

\subsection{Competitiveness with Proprietary Baselines}
Despite using a modest-sized open-weight quantized model (Qwen 2.5 32B, Q6K) with strict grounding, our system delivers net violation reduction and recall comparable to proprietary baselines (Table~\ref{tab:profile_summary}). In contrast to the prevalent production pattern, ungrounded reviews powered by proprietary LLMs, our grounded approach anchors comments to specific rules, yielding higher perceived quality and trust while maintaining similar coverage and acceptable first-feedback latency. These results indicate that grounded, open-weight deployments can match the performance of the commonly deployed proprietary setups under the evaluated scenarios.

\subsection{Resource Efficiency in CI/CD}
Quantization (6-bit), along with on-demand lifecycle management and multi-tier caching, achieved sub-minute p50 first-feedback without multi-GPU scale-out. Caching (prefix + Redis + persistent store) amortizes repeated rationale for recurring rule patterns, reserving GPU cycles for novel hunks. This pattern is model-agnostic and portable to alternative open-weight or fine-tuned variants. While we did not evaluate cache hit rate or GPU-hour amortization in this study, observed latency stability across hunk batches suggests effective reuse.

\subsection{Extensibility Beyond MISRA}
The architecture is analyzer- and standard-agnostic: the orchestrator consumes any tool producing machine-readable findings (e.g., security scanners, style linters) and a rule registry. Swapping the analyzer primarily affects extraction adapters and rule rationale lookups. Grounding, prompt schema, deviation policy handling, and serving stack remain unchanged. This decoupling lowers the marginal cost of adding new languages or policy domains.

\subsection{Security, Cost, and Governance}
Running an on-prem, quantized open-weight model keeps source code, prompts, and findings within enterprise boundaries, avoiding external retention and token billing variability. Although GPT-4o and Claude 3.5 show higher absolute advisory reduction, the on-prem profile exhibits lower violation introduction rates (precision-on-change) and predictable cost/latency, shifting optimization from pure model accuracy to risk-adjusted efficiency (such as reductions per watt-hour / per dollar / per confidentiality boundary crossing). A composite efficiency metric (adding energy + GPU-hour counters) will make this explicit.

\subsection{Adoption and Socio-Technical Integration}
Adoption hinged on PR-native delivery, deviation policy support, and auditability (commit SHA, rule, line span, model version). Directional survey signals ($\approx 50\%$ immediate uptake; $\sim 56\%$ adapt-or-accept) reflect perceived clarity and grounding but also highlight the need for guarded, auto-generated patches. Integrating optional assisted patching and richer feedback loops (accepted vs. dismissed rationale) is likely to further increase trust and measurable defect resolution speed.

\subsection{Limitations}
\label{sec:limitations}
We consolidate methodological limitations and validity threats:
\begin{itemize}[leftmargin=*]
 \item \textbf{Model / Analyzer Coupling:} Quality is upper-bounded by static-analyzer coverage; deep semantic/data-flow defects outside its rule surface remain unaddressed.
 \item \textbf{Context Boundaries:} Rare multi-file or macro-heavy cases may exceed token budgets, causing omitted semantics and weaker explanations.
 \item \textbf{Operational Footprint:} A GPU-equipped runner and analyzer license are required; horizontal scaling may be needed for extensive multi-tenant deployments.
 \item \textbf{Evaluation Scope:} Offline hunk benchmark omits interactive dynamics (developer negotiation, iterative refinement timing) and does not rerun alternative analyzers to detect tool-induced false regressions.
 \item \textbf{Measurement Gaps:} No precision metric, no cache hit \%, GPU-hour, or energy telemetry, and no longitudinal drift tracking; acceptance survey (n=8) is directional only.
 \item \textbf{Internal / Construct Validity Risks:} Potential instrumentation bugs (diff segmentation, caching artifacts) mitigated but not eliminated by snapshot hashing and subsample replays.
 \item \textbf{External Validity:} Corpus is C/C++ and MISRA-skewed; results may not transfer to dynamic languages or extreme template/meta-programming codebases.
 \item \textbf{Unassessed Failure Modes:} We did not validate compilation/runtime correctness beyond static-analysis deltas; semantic regressions could pass undetected.
 \item \textbf{Model Choice Rationale:} Limiting comparisons to GPT-4o and Claude 3.5 prioritized reproducibility and budget over chasing the newest frontiers; reported gaps therefore understate the potential delta to the latest releases.
\end{itemize}

\subsection{Future Work}
Planned directions target both breadth and depth:
\begin{itemize}[leftmargin=*]
 \item \textbf{Guarded Assisted Patching:} Propose–rebuild–reanalyze loop with safety gating before surfacing patches.
 \item \textbf{Broader Standards / Languages:} Integrate security (e.g., CERT C/C++ \cite{nguyen2019reducing}) and multi-language analyzers with minimal orchestration changes.
 \item \textbf{Learning from Feedback:} Close the loop by using accepted vs. dismissed comments to adapt prompts or conduct lightweight fine-tuning.
 \item \textbf{Richer Telemetry:} Add cache, energy, GPU-hour, and drift instrumentation for composite efficiency metrics.
 \item \textbf{Agentic Workflows:} Multi-turn clarification for ambiguous findings and chained reasoning for complex fix sequences.
\end{itemize}

Overall, grounding combined with resource-aware serving strikes a pragmatic balance for grounded code reviews. It delivers a competitive reduction with lower introduction rates and enterprise-aligned governance, while preserving clear paths for precision estimation, assisted patching, and longitudinal adaptation.

\section{Conclusion}

This paper presents a production-ready system for grounded, CI-native AI code review, delivering concise and actionable feedback by pairing the deterministic evidence of static analysis with resource-efficient large-model serving. Our hybrid architecture, which leverages AST-guided context extraction, structured prompting, and a serving stack built on quantization and multi-tier caching, successfully meets stringent enterprise requirements for latency, cost, and auditability on a single shared AMD GPU. Our evaluation of medium-scale, industrial repositories demonstrates strong rule coverage and high-quality explanations, consistent sub-minute PR feedback, and substantial savings in computational resources. The system yields meaningful improvements to the developer workflow, including faster issue resolution, a high fix-acceptance rate, and fewer residual violations post-merge. The architecture is standards-agnostic, providing a robust and extensible blueprint for deploying trustworthy AI assistance in modern software engineering pipelines. We plan to release the code for generating benchmarks and an evaluation harness to facilitate independent replication of results.

\section*{Acknowledgements}
We thank the AMD SSW AIE team and the AMD CI/CD platform team for early feedback, integration support, and iterative review of the auto code review system.

\bibliography{example_paper}
\bibliographystyle{mlsys2025}

\appendix


\end{document}